\documentclass{article}
\usepackage{spconf,amsmath,graphicx}
\usepackage{float}
\usepackage{amssymb}

\title{WEAKENING THE DETECTING CAPABILITY OF CNN-BASED STEGANALYSIS}
\name{Sai Ma, Qingxiao Guan$^*$, Xianfeng Zhao, Yaqi Liu }
\address{Institute of Information Engineering, Chinese Academy of Sciences, Beijing 100093, China}
\begin{document}
%
\maketitle
\begin{abstract}
Recently, the application of deep learning in steganalysis has drawn many researchers' attention. Most of the proposed steganalytic deep learning models are derived from neural networks applied in computer vision. These kinds of neural networks have distinguished performance. However, all these kinds of back-propagation based neural networks may be cheated by forging input named the adversarial example. In this paper we propose a method to generate steganographic adversarial example in order to enhance the steganographic security of existing algorithms. These adversarial examples can increase the detection error of steganalytic CNN. The experiments prove the effectiveness of the proposed method.

\end{abstract}
\section{Introduction}
\label{sec:intro}
Steganography is the science and the art of covert communication via digital media. Accompanied by steganography, steganalysis is the methodology to detect the trace of secret message in the media. As the steganographic algorithms embed the message in the media via slightly modifying the image elements, steganalysis applies statistical methods to catch the trace of these kinds of modifications. To decrease the possibility of being detected by steganalyzer, the modern steganography focus on adaptive schemes.

Adaptive steganography is the scheme that embeds the message according to the content of media. The scheme modifies the elements in complicated area of the image, so the modifications can be concealed by the local content. To detect adaptive steganography effectively, steganalytic method equipped with Rich Model \cite{rmTIFS} is invented. Rich Model is the steganalytic feature extraction scheme which can capture the high order statistical feature of the stego image.

Recently, researchers start to apply deep learning for steganalysis. These years have seen a great progress on it. The first work was proposed in \cite{dlstegFirst}. Although the performance is not outstanding, it is an early attempt. Later, the work that can nearly challenge the Rich Model was proposed in \cite{dlstegQian}. In 2015, Xu et al. proposed a neural network \cite{dlstegXu} whose performance is better than Rich Model. With these beneficial exploration, deep learning based steganalysis has become one of the mainstream of research \cite{dlstegTIFS}. Comparing with handy-crafted-designed steganalysis, the advantage of deep learning based steganalysis is its "end-to-end" framework. Despite the structure of the network and super parameters for training is needed to be manually determined, the parameters of the neural network is optimized by machine during training.

With development of deep learning, researchers found that the neural network which relies on back-propagation (BP) has an interesting property \cite{adverSze}\cite{adverGoodf}. Such property is that the BP-based neural network may mis-classify the input data when the data is added a "noise" map generated from gradient feature map of the neural network. This kind of data for "cheating" the BP-based neural network is named as the adversarial example.

In this paper we propose a novel method to enhance the security of the existing adaptive steganographic algorithms against the deep-learning-based steganalysis. Given a optimized steganalytic CNN model, we can generate the adversarial examples of the input images. The adversarial example can "cheat" the steganalytic CNN (Convolutional Neural Network), so as to decrease the possibility of being detected. The proposed method exploits the gradient feature map to determine the flipping direction of pixels while embedding, the flipping is equivalent to $\pm1$. This operation perturbs the classification of steganalyzer and makes the classification result lean to cover. The experiments show that the steganographic adversarial example can significantly increase the detection error rate of the steganalyzer. We note that the proposed method is a forging method to protect the stego image from being detected, which is different from recently prevailing Generative Adversarial Network (GAN). GAN based method is to generate stego image \cite{ganShi} or distortion function \cite{ganSPL}, which needs to train a new generative neural networks. The proposed method in this paper does not build new neural network, instead, it generates adversarial data from steganalyzer but does not update the parameters of the steganalyzer. Our goal is to enhance the security of existing steganographic algorithms.

\section{Preliminaries}
\label{sec:prelimi}

\subsection{notations}
\label{ssec:nota}
A digital image can be represented in either a matrix or a vector. Matrix is noted with capital boldface $\textbf{X}$, its element in t$(i, j)$ is noted as $X_{ij}$. Vector is noted with lower case boldface $\textbf{v}$, the element in the $i$-th dimension is noted as $v_{i}$.

In this paper, we focus on the spatial gray-scale image. Therefore, a pixel $x_{i} \in \{0,...,255\}$.  The steganographic distortion of pixel $x_{i}$ is noted as $\rho_{i}$ and $\rho_{i} \geqslant 0, \rho_{i} \in \mathbb{R}$.

\subsection{DM framework}
\label{ssec:dm}

The adaptive steganographic framework has become the fundamental infrastructure in recent years. The framework can be also abbreviated to DM (distortion-minimization) framework \cite{hugo}\cite{wow}\cite{suniward}\cite{hill}. The DM framework consists of distortion function and adaptive coding method.

\subsubsection{distortion function}
\label{sssec:disto_func}
Given a image, steganographer slightly changes some pixels to embed the secret message in the image. Due to the modification, the statistical perturbation is introduced into the image inevitably. The steganalytic schemes are designed to detect this kind of perturbation so as to distinguish the normal image and the stego image. The target of DM framework is to minimize the statistical perturbation of the steganographic modification. However, it is difficult to model the statistical effect of stego embedding because of the data's high dimension and correlation of pixels. To quantify the effect of the modification, the steganographic distortion function is proposed. The distortion function is defined as the metric of statistical perturbation caused by steganographic modification. To be a practical solution of DM framework, it is considered to be additive. The cost value of a pixel is heuristically calculated based on neighbors. From distortion function, every pixel is assigned a profile as the cost of being changed. The profile in the complex region is lower than that in the smooth region. Intuitively, such property of distortion function makes the modifications gathering in the complex regions.

\subsubsection{adaptive coding method}
\label{sssec:ada_coding}
As cost values are assigned to the pixels, next target is to embed the secret message in the image while minimizing the sum of cost values. The state-of-the-art coding method is called STC \cite{stcTIFS} (Syndrome-Trellis Code). Next, we will briefly introduce STC.

We note that $\textbf{x} \in \{0, 1\}^n$ is the LSB vector of the cover pixels and $\textbf{y} \in \{0, 1\}^n$ is the LSB vector of the cover pixels, $\textbf{m} \in \{0, 1\}^l$ is the secret message sequence to be embed, $\textbf{H} \in \{0, 1\}^{l \times n}$ is a binary matrix. The steganographic embedding task is modifying $\textbf{x}$ to $\textbf{y}$. The modification pattern $\textbf{s}$, which $\textbf{s}=\textbf{x} - \textbf{y}$. The task is an optimizing problem:
\begin{equation}\label{eq:cost_sum}
\begin{aligned}
\arg \min \sum_{i = 1}^n \rho_i \cdot s_i \\
s.t. \textbf{H} \times \textbf{y} = \textbf{m},
\end{aligned}
\end{equation}
where $\rho_i$ is $i$-th pixel's cost value and $s_i$ is $i$-th element of modification pattern $\textbf{s}$. STC is derived from Viterbi algorithm and it can solve the problem (\ref{eq:cost_sum}) near-optimally.

In this paper we apply single-layered STC as the embedding method. Single-layered STC is often applied for LSBR (LSB replacement) embedding, which is changing the LSB of selected pixel. Besides LSBR, there are several methods to change the LSB. In this paper, we use $\pm1$ to flip LSB and use single-layered STC to determine the pixels to be changed.

\section{STEGANOGRAPHIC ADVERSARIAL EXAMPLE}
\label{sec:stego_adver}

\subsection{adversarial example}
\label{ssec:adver}
In 2014, Szegedy et al. found an intriguing property of neural networks that the BP-based neural network will mis-classify the image when the input data is added a well-designed noise map derived from back-propagated gradient residual \cite{adverSze}. The perturbation is invisible but is sensitive to the neural network. Such kinds of input data that can cause mis-classification is called the adversarial example.

Although neural networks have many variations of architectures, all of them is the mapping operation. In the task of steganalysis, the mapping function is denoted as $F: \mathbb{R}^{512 \times 512} \rightarrow \{[1, 0], [0, 1]\}$. The input of neural network is a image and the output is a label vector $[cover, stego]$. The output result is determined by the probabilities of 2 categories. We note the probability of cover image as $p_c$ and probability of stego image as $p_s$. $p_c$ and $p_s$ satisfy the relation:
\begin{equation}\label{eq:probs}
p_c = 1 - p_s
\end{equation}

To make further explanation, let $p_c = F_c(\textbf{X})$. For a trained model, $F_c$ is differentiable almost everywhere. Given an image $\textbf{X}$, we can compute the gradient matrix of $F_c$. The gradient matrix of $F_c$ is denoted as $\textbf{G}$:
\begin{equation}\label{eq:grad_mat}
\textbf{G} = \nabla F_c(\textbf{X}) =
\begin{aligned}
\begin{bmatrix}
  \frac{\partial F_c(\textbf{X})}{\partial X_{1,1}} & \cdots & \frac{\partial F_c(\textbf{X})}{\partial X_{1,n}} \\
  \vdots & \frac{\partial F_c(\textbf{X})}{\partial X_{i,j}} & \vdots \\
  \frac{\partial F_c(\textbf{X})}{\partial X_{m,1}} & \cdots & \frac{\partial F_c(\textbf{X})}{\partial X_{m,n}} \\
\end{bmatrix}
\end{aligned}
\end{equation}
the element in $(i, j)$ is the gradient value of the pixel $X_{ij}$. When $G_{ij} > 0$, within the local neighborhood of $X_{ij}$, $F_c(\textbf{X})$ is increasing. On the contrary, $F_c(\textbf{X})$ decreases within the local neighborhood when $G_{ij} < 0$.

From equation (\ref{eq:probs}) we know:
\begin{equation}\label{eq:softmax}
\nabla p_c = - \nabla p_s,
\end{equation}
which indicates that these two probabilities shift in opposite directions. To a steganographer, he can use the back-propagated gradient feature map to control the classification result, in order to cheat the steganalyzer.

\subsection{fast gradient sign method}
\label{ssec:fgsm}
In 2015, Goodfellow et al. proposed a method to generate the adversarial example, which is called Fast Gradient Sign Method (FGSM) \cite{adverGoodf}. The method is simple to implement and has a remarkable effect on cheating the deep learning model.

To empathize the variables of the function, the mapping function of the model is denoted as $F_c(\textbf{x}, \boldsymbol{\theta})$. In the function, $\textbf{x}$ is the input data, $\boldsymbol{\theta}$ is the parameters of the model.

To generate a adversarial example, the easiest solution is adding the CNN's gradient feature map to the input data. However, for a optimized model, the back-propagated signal is too weak to have a significant effect. As the sign of the gradient value ($\pm$) indicates the changing trend of the function, we apply the signum function ($y=sgn(x)$) to the gradient value instead of its original value. The perturbation $\boldsymbol{\eta}$ is computed as follows:
\begin{equation}\label{eq:fgsm}
\boldsymbol{\eta} = \epsilon \cdot sgn(\nabla_{\textbf{x}}F_c(\textbf{x}, \boldsymbol{\theta}))
\end{equation}
where $\epsilon$ is the magnitude of the perturbation signal. In theoretical analysis, $\epsilon \in \mathbb{R}^+$. However, in application scenario, the CNN's input data is usually 8-bit integer, so we set $\epsilon = 1$.

After generating perturbing feature map, we can add it to the original input data:
\begin{equation}\label{eq:add_adver}
\tilde{\textbf{x}} = \textbf{x} + \boldsymbol{\eta},
\end{equation}
$\tilde{\textbf{x}}$ is the adversarial example of $\textbf{x}$. In \cite{adverGoodf}, the authors choose GoogLeNet \cite{googlenet} to test the method. The adversarial example is mis-classified by the model and has a very high confidence (99.3\%). Therefore, FGSM is proved to be an effective method. In this paper, we generate $\boldsymbol{\eta}$ of the cover image and add it to the stego image. The details will be discussed in the next section.

\subsection{steganographic adversarial example}
\label{ssec:stego_adver}
Inspired by \cite{adverSze}\cite{adverGoodf}, we proposed a method to enhance the security of DM steganographic algorithms against deep learning based steganalysis.

In our method, we apply single-layered STC to embed the message. As mentioned in section \ref{sssec:ada_coding}, we use single-layered STC to determine the pixels to be changed. Despite a small number of wet pixels (0 or 255), both +1 and -1 are suitable for changing the LSB. Therefore, we can determine the flipping direction of the pixel to be modified according to the sign of its gradient residual.

\begin{figure}[htb]
  \centering
  \includegraphics[width=0.68\columnwidth]{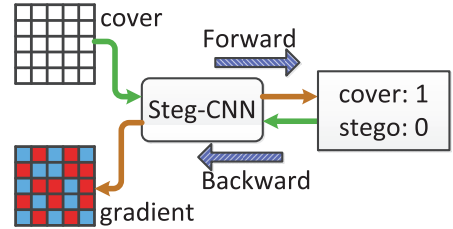}
  \caption{Generating cover's gradient feature map}
  \label{fig:gen_grad}
\end{figure}

Our task is to cheat the steganalytic neural network. The first step is generating the gradient feature map of the cover. Given a steganalytic CNN model and a cover image, we first input the image into CNN, and get the classifying result. Next, we conduct the back-propagation to get the gradient feature map of the cover. In the case of that cover is classified as stego, we force the output label to be cover, in order to make sure that gradient feature map tends to cover. Step 1 is illustrated in Fig. \ref{fig:gen_grad}. In Fig. \ref{fig:gen_grad}, the $5\times5$ blank matrix on the left top represents the cover image. The cover is the input of the steganalyzer (block in the middle, labelled with "Steg-CNN"). The output (right) is the classifiction result, which is a label vector. As the input is cover, so the output label vector is $[1, 0]$. Then we conduct back-propagation. The gradient feature map is on the left bottom. In the gradient feature map matrix, red grid is +1 and blue gird is -1.

\begin{figure}[htb]
  \centering
  \includegraphics[width=0.68\columnwidth]{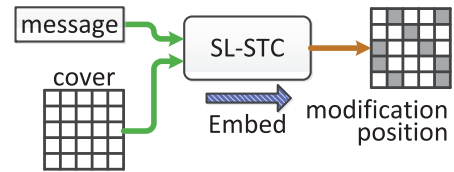}
  \caption{Determining modification positions}
  \label{fig:sl_stc}
\end{figure}

Step 2 is illustrated in Fig. \ref{fig:sl_stc}, which is determining the positions to be changed in the cover image. This step is a typical DM steganographic embedding operation. In this step we apply single-layered STC (abbreviated to SL-STC in Fig. \ref{fig:sl_stc}) to determine the pixels to be changed. In the figure, input is the cover (left) and output (right) is the matrix that implies the pixels to be changed. The gray-colored grid in the matrix means that the pixel is to be modified and white grid means the pixel is unchanged.

\begin{figure}[htb]
  \centering
  \includegraphics[width=0.68\columnwidth]{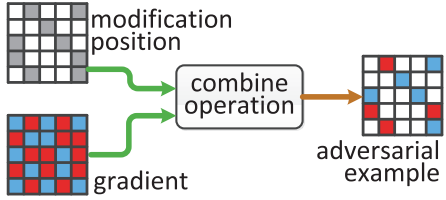}
  \caption{Generating adversarial example}
  \label{fig:combine}
\end{figure}

Step 3 is generating the adversarial example. In Step 1 we get the gradient feature map which can make the stego image look like the cover image. In Step 2 we know the pixels to be modified. The task of Step 3 is to actually modify the cover image according to the sign of gradient residual. The modification of each pixel is either +1 or -1. We note that if the pixel is a wet pixel, which means that its value is 0 or 255, its flipping direction is fixed to be +1 or -1, respectively. In Fig. \ref{fig:combine}, the output matrix (right) is the adversarial stego image.

\section{EXPERIMENT}
\label{sec:exp}
In this section, we test the performance of our method. We choose Xu's CNN and maxSRM + ensemble classifier \cite{maxsrm}\cite{ensemble} as steganalyzers. We apply average detection error rate $\bar{P_E}$ of the steganalyzer to evaluate the performance of the proposed method. $\bar{P_E}$ is calculated as follows:
\begin{equation}\label{eq:aver_pe}
\bar{P_E} = \frac{1}{2}(P_{FA}+P_{MD})
\end{equation}
where $P_{FA}$ is the false alarm rate and $P_{MD}$ is the mis-detection rate.

We use BOSSbase (10000 images) as the image base. We choose five payload rates to generate stego images, which are 0.05, 0.1, 0.2, 0.3, 0.4. We choose S-UNIWARD and HILL to generate stego images. Under each payload rate, we randomly choose 5000 images in BOSSbase as cover images and generate their corresponding stego images via single-layered STC. Therefore, for each payload rate, there is a training set with 5000 cover/stego pairs. With a training set, we train a CNN model and an ensemble classifier. The rest 5000 images are used to generate testing sets. Firstly, being as the same as generating training set, we generate 5000 cover/stego pairs as a testing set under each payload rate. Then, for every stego image in the testing set, we generate an adversarial stego image from its corresponding payload rate's CNN model via our method, so we have an another testing set with 5000 cover/adversarial pairs. We note that these two testing sets (5000 cover/stego and 5000 cover/adversarial) are not used to train the steganalyzers. The original model of Xu's CNN uses five identical neural networks. For convenience, we use one neural network here.

\begin{figure}[htb]
  \centering
  \includegraphics[width=0.7\columnwidth]{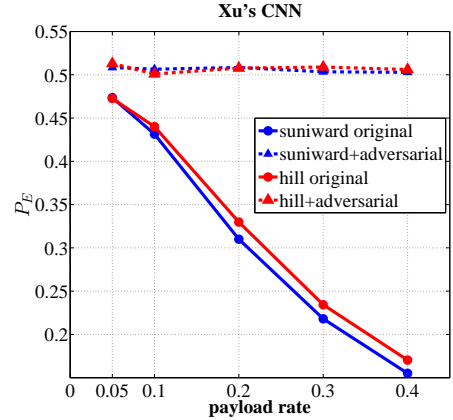}
  \caption{Results on Xu's CNN}
  \label{fig:exp1}
\end{figure}

\begin{figure}[htb]
  \centering
  \includegraphics[width=0.7\columnwidth]{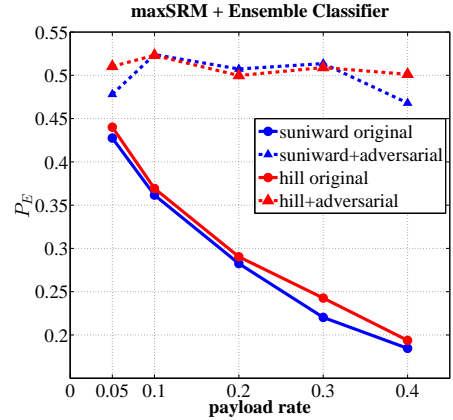}
  \caption{Results on maxSRM}
  \label{fig:exp2}
\end{figure}

Fig. \ref{fig:exp1} is the experiment result on Xu's CNN and Fig. \ref{fig:exp2} is the result on maxSRM + ensemble classifier. The data in circle is the result of "original stegos" (without adversarial examples) and data in triangle is the result of adversarial examples. From the results of two experiments we can find that the proposed method has good performance both in CNN based-method and in feature-based method. It is proved that our method is effective to enhance the security of existing steganographic algorithms.

\section{CONCLUSION}
\label{sec:conclu}
In this paper we propose a novel method to enhance the security of steganographic algorithm towards the attack of deep learning based steganalysis. As the BP-based neural network can be cheated by forging data named adversarial examples, we apply this forging method in steganography. The experiments prove the effectiveness of our method.

Encouraged by the result of experiments, we will step forward in the future work. In this paper we generate the adversarial examples via single-layered STC in spatial domain. In the future, we will design the adversarial method applied on double-layered STC.

Finally, we need to emphasize that the adversarial example method applied in this paper can only deceive the steganalyzer trained with non-adversarial example. To enhance the steganographic security, we need to further explore the adversarial method. This is also a part of our future work.

%


\vfill\pagebreak

\bibliographystyle{IEEEbib}
\bibliography{mylib}

\end{document}